\documentclass[singlecolumn,amssymb, nobibnotes, show---pacs, superscriptaddress, aps, prd]{revtex4}
\usepackage{graphicx,amsmath,amssymb,color}

\usepackage{soul}

\begin{document}
\title{Colliding-probe bi-atomic magnetometers via energy circulation:\\ 
	Breaking symmetry-enforced magneto-optical rotation blockade}

\author{L. Deng}
\affiliation{Center for Optics Research and Engineering (CORE), Shandong University (Qingdao), China}
\email{lu.deng@email.sdu.edu.cn}

\date{\today}

\begin{abstract}
	We have developed an inelastic wave scattering based colliding-probe bi-atomic magnetometer theory. We show a propagation growth blockade in single-probe-based magnetic field sensing schemes, revealing the root-cause of strong suppression of nonlinear magneto-optical rotation effect (NMORE) in single-probe-based atomic magnetometers. We further show, both experimentally and theoretically, a colliding-probe bi-atomic magnetometer that lifts this NMORE blockade. The directional energy-circulation in this new atomic magnetometry technique results in more than two orders of magnitude increase in NMORE signal as well as greater than 6-dB increase of magnetic field detection sensitivity. The new technique may have broad applications in photon gates and switching operations.
\end{abstract}

\maketitle
	
	\section{Introduction}
	Atomic magnetometers (AMs) describe a class of magnetic field sensing devices that operate based on the principles of the magneto-optical rotation effect. The unique spectral density of lasers has led to demonstration of the dependency of the rotation effect on laser power, for which the nonlinear magneto-optical rotation effect (NMORE) was named \cite{r1,r2}. For the past 60 years, AMs have primarily been a single-probe laser technique where a linearly polarized electromagnetic field and an $F=1$ atomic specie form the core of the technique \cite{r3,r4,r5}. Over the years pioneering researches and innovations \cite{r6,r7,r8,r9,r10,r11,r12,r13,r14} have contributed to the advancement of this vibrant field. However, to date there has been no study that raises the question of why NMORE$-$a ``nonlinear effect"$-$does not have the characteristic nonlinear propagation dependency widely seen in nonlinear optics \cite{r21}. Here, we show experimentally and theoretically that the root cause of this ``nonlinear effect without nonlinear propagation characteristics" is an NMORE blockade. It is enacted by symmetries in atomic population distribution and transition rates, as well as cross-component energy conservation in the singe-probe excited $F=1$ system. As a result, no significant energy flow is permitted and the NMORE is strongly suppressed. Furthermore, we demonstrate a colliding-probe bi-atomic magnetometry technique where the introduction of a second probe field can effectively lift this NMORE blockade. This results in a significant directional energy circulation and a giant NMORE signal-to-noise ratio (SNR) enhancement, as well as increased magnetic field sensitivity. Remarkably, all experimental observations can be well explained by this inelastic-wave-scattering-based \cite{r15,r16} colliding-probe bi-AM theory, including the predicted giant NMORE signal-to-noise ratio (SNR) enhancement, magnetic field sensitivity increase, and NMORE polarization cross angle dependency \cite{r17,r18,r19}.
	
	\section{Experimental observations and Theoretical framework}
	\noindent
	Figures 1(a) shows a representative NMORE fast-Fourier-transform (FFT) spectrum (red) using the colliding-probe bi-AM technique where a 10-nT magnetic field is modulated at 20 Hz. For comparison, we also show the corresponding single-probe AM NMORE FFT spectrum (blue) under the same experimental conditions (i.e., just turned the second probe off). With a 1-cm vapor cell at human body temperature we routinely observe more than two orders of magnitude NMORE signal enhancement using the colliding-probe bi-AM technique and very weak probe intensities. Without the second probe, no NMORE signal can be seen (blue trace). However, when the second probe is present, a giant NMORE signal (typically $>$20-dB) can be easily detected. Figure 1(b) shows the cross-polarization angular effect observed with the colliding-probe bi-AM.
	\begin{figure}[htb]
		\centering
		\includegraphics[width=14 cm]{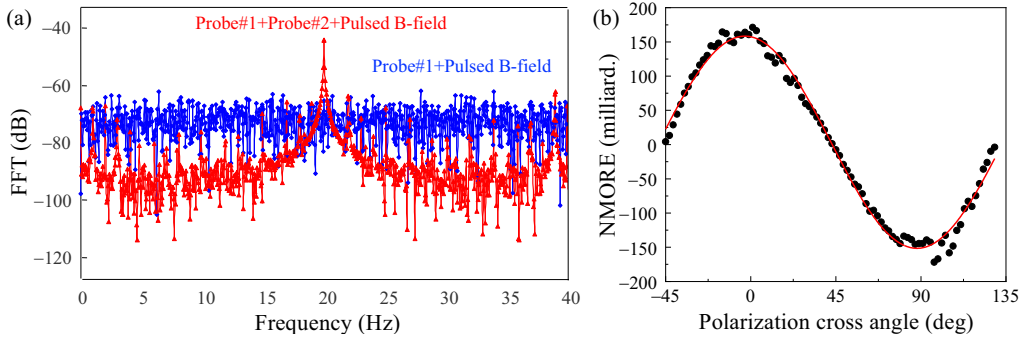}
		\caption{{\bf Giant NMORE signal enhancement by colliding-probe bi-AM technique.} (a) Representative NMORE FFT spectra: Colliding-probe bi-AM (red trace) vs. the conventional single-probe AM (blue trace) under the same experimental conditions. (b) Cross-polarization angular effect of the colliding-probe bi-AM. Typically, lasers are detuned by a few GHz away from relevant one-photon resonances (see Fig. 2). }
	\end{figure}
	\vskip 10pt
	\noindent The experimental observations shown in Figs. 1(a) and 1(b) can be qualitatively explained by an inelastic-wave-scattering theoretical framework based colliding-probe bi-AM theory \cite{r16,r17}. Here, we consider a four-state atomic system, depicted in Fig. 2(a), where the atomic state $|j\rangle$ has energy $\hbar\omega_j$ ($j=1,...,4$) and the lower three states form an $F=1$ system. We assume that the probe field $\mathbf{E}_{p1}$ (frequency $\omega_{p1}$) is polarized along the $\hat{x}$-axis and propagates along the $\hat{z}$-axis. Its $\sigma^{(\pm)}$ components couple independently the $|1\rangle\Leftrightarrow|2\rangle$ and $|3\rangle\Leftrightarrow|2\rangle$ transitions with a large one-photon detuning $\delta_{p1}=\delta_2=\omega_{p1}-[\omega_2-(\omega_1+\omega_3)/2]$. Initially the population is equally shared by the two ground states $|F=1, m_F=\pm 1\rangle$, therefore two opposite two-photon transitions between states $|1\rangle$ and $|3\rangle$ with a two-photon detuning $2\delta_B$ are simultaneously established. Here, the Zeeman frequency shift $\delta_B=g\mu_0 B$ in the axial magnetic field $B=B_z$ is defined with respect to the mid-point between the two equally but oppositely shifted Zeeman levels $|1\rangle=|m_F=-1\rangle$ and $|3\rangle=|m_F=+1\rangle$. A second probe field $\mathbf{E}_{p2}$ (frequency $\omega_{p2}$) propagates along the $-\hat{z}$-axis but has its linear polarization at an angle $\theta_0$ with respect to the $\hat{x}$-axis. Its
	$\sigma^{(\pm)}$ components independently couple the $|3\rangle\Leftrightarrow|4\rangle$ and $|1\rangle\Leftrightarrow|4\rangle$ transitions with a large one-photon detuning $\delta_{p2}=\delta_4=\omega_{p2}-[\omega_4-(\omega_1+\omega_3)/2]$, forming again two two-photon transitions between states  $|1\rangle$ and $|3\rangle$ with the same two-photon detuning $2\delta_{B}$. More precisely, there are two excitation channels (each contains two {\it competing} two-photon transitions. i.e., $|1\rangle\Leftarrow|3\rangle$ and $|1\rangle\Rightarrow|3\rangle$) that share the $\bf same$ equally-populated states $|1\rangle$ and $|3\rangle$, creating a mutually influencing ground-state Zeeman coherence. We stress that the underlying physics of this new colliding-probe NMORE technique is inelastic wave scattering which has absolutely {\bf no} resemblance to the usual elastic four-wave mixing or electromagnetically induced transparency based processes (see discussion later). A close analogy of this inelastic wave scattering process is the light scattering in an equally populated ground state spinors system of an $F=1$ atomic Bose-Einstein condensate. 
	\begin{figure}[htb]
		\centering
		\includegraphics[width=14 cm]{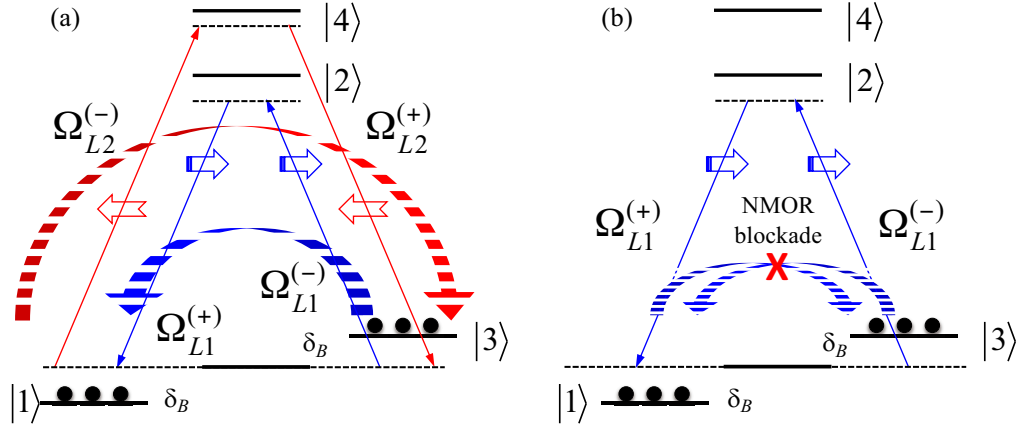}
		\caption{{\bf Energy level diagrams.} (a) Energy level diagram of Colliding-probe bi-AM showing energy circulation which results in a significant NMORE SNR enhancement. Probe $\mathbf{E}_{p1}$ couples $|5S_{1/2},F=1\rangle\rightarrow|5P_{1/2},F^{'}=1\rangle$ transition with a detuning of $-$5GHz. Probe $\mathbf{E}_{p2}$ couples $|5S_{1/2},F=1\rangle\rightarrow|5P_{3/2},F^{''}=2\rangle$ transition with a detuning of $-$3GHz. Hollow arrows represent the propagation directions of the relevant fields. (b) Energy level diagram of the conventional single-probe AM showing an NMORE blockade (two wide dashed arrows with the red X).}
	\end{figure}
	\vskip 10pt
	\noindent Under the electric-dipole approximation, the system interaction Hamiltonian is given as
	\begin{eqnarray}
	\frac{\hat{H}}{\hbar}\!=\!\sum_{j=1}^4\delta_j|j\rangle\langle j|\!+\!\sum_{m=2,4}\!\left[\Omega_{m1}|m\rangle\langle 1|\!+\!\Omega_{m3}|m\rangle\langle 3|\!+\!{\rm c.c}\right],\:
	\end{eqnarray}
	where $\delta_j$ is the laser detuning from state $|j\rangle$. The total electric field is given by $\mathbf{E}\!=\!\sum_{a}\mathbf{E}_{a}$ where
	$\mathbf{E}_{a}\!=\!\left(\mathbf{\hat{e}}_{+}{\cal E}_{a}^{(+)}\!+\!\mathbf{\hat{e}}_{-}{\cal E}_{a}^{(-)}\right){\rm e}^{i\theta_{a}}\!+\!{\rm c.c.}$ and $\theta_{a}\!=\! \mathbf{k}_{a}\!\cdot\mathbf{r}-\omega_{a}t$ with $k_{a}\!=\omega_{a}/c$ being the wavevector of the field $\mathbf{E}_{a}$. Here, $a\!=\!p1,p2$, and $\mathbf{k}_{a}\!=\!\pm\hat{z}k_{a}$ where the $\pm$ sign distinguish the propagation directions of the two probe fields. Correspondingly, $\Omega_{21}=\Omega_{p1}^{(+)}=D_{21}{\cal E}_{p1}^{(+)}/(2\hbar)$ and $\Omega_{23}=\Omega_{p1}^{(-)}=D_{23}{\cal E}_{p1}^{(-)}/(2\hbar)$, $\Omega_{41}=\Omega_{p2}^{(-)}=D_{41}{\cal E}_{p2}^{(-)}/(2\hbar)$ and $\Omega_{43}=\Omega_{p2}^{(+)}=D_{43}{\cal E}_{p2}^{(+)}/(2\hbar)$. $D_{nm}=\langle n|\hat{D}|m\rangle$ is the transition matrix element of the dipole operator $\hat{D}$.
	
	\vskip 10pt
	\noindent
	Under the rotating wave approximation the Schrodinger equations describing wave-function amplitudes are,
	\begin{subequations}
		\begin{align}
			&\dot{A}_{1}-i\delta_BA_1=i\Omega_{12}A_2+i\Omega_{14}A_4-\gamma A_1,\\
			&\dot{A}_{3}+i\delta_BA_3=i\Omega_{32}A_2+i\Omega_{34}A_4-\gamma A_3,\\
			&\dot{A}_{2}-i\delta_{p1}A_2=i\Omega_{21}A_1+i\Omega_{23}A_3-\Gamma A_2,\\
			&\dot{A}_{4}-i\delta_{p2}A_4=i\Omega_{41}A_1+i\Omega_{43}A_3-\Gamma A_4,
		\end{align}
	\end{subequations}
	where for simplicity we have expressed the decay rates of the ground and excited states as $\gamma$ and $\Gamma$, respectively.
	
	\vskip 10pt
	\noindent
	The third-order perturbation calculation proceeds as follows \cite{r21}. First, with large one-photon detunings we obtain adiabatic solutions for $A_2$ and $A_4$. We then seek corrections to steady-state solutions of ground states up to the third-order of probe fields. Finally, the polarization source terms for Maxwell equations of circularly-polarized probe components $\Omega_{p1}^{(\pm)}$ are constructed and the properties of these differential equations are analyzed.
	
	\vskip 10pt
	\noindent
	Applying the above prescribed perturbation method we obtain for the $\Omega_{p1}^{(+)}$ component,
	\begin{eqnarray}
	\rho_{21}^{(3)}&=\frac{S_{p1}^{(-)}}{2\Gamma(1-id_{p1})}\left\{\frac{\Omega_{p1}^{(+)}}{(1+d_{p1}^2)}\left(\frac{1+id_{p1}}{\beta_{-}+i\Gamma_{-}}-\frac{1-id_{p1}}{\beta_{+}+i\Gamma_{+}}\right)+\frac{\Omega_{p2}^{(-)}}{(1+d_{p2}^2)}\left(\frac{1+id_{p2}}{\beta_{-}+i\Gamma_{-}}-\frac{1-id_{p2}}{\beta_{+}+i\Gamma_{+}}\right)\right\},
	\end{eqnarray}
	where $d_{p1}\!=\!\delta_{p1}/\Gamma$, $d_{p2}\!=\!\delta_{p2}/\Gamma$, 
	$d_B\!=\!\delta_B/\gamma$, $\beta_{\pm}\!=\!-d_B\pm B_{\pm}$, and $\Gamma_{\pm}\!=1\!+H_{\pm}$. Here,  $B_{\pm}\!=\!d_{p1}S_{p1}^{(\pm)}\!/(1+d_{p1}^2)+ d_{p2}S_{p2}^{(\mp)}\!/(1+d_{p2}^2)$ and $H_{\pm}\!=\!S_{p1}^{(\pm)}\!/(1\!+\!d_{p1}^2)\!+\!S_{p2}^{(\mp)}\!/(1\!+\!d_{p2}^2)$ are the total light induced frequency shift and resonance broadening. In addition, the channel-specific two-photon saturation parameters are defined as $S_{a}^{(\pm)}=|\Omega_{a}^{(\pm)}|^2/\gamma\Gamma$. For large one-photon detunings, $B_{\pm}$ and $H_{\pm}$ are negligible, and $\beta_{-}\!=\!\beta_{+}$, $\Gamma_{-}\!=\!\Gamma_{+}$. This is one of the key advantages of the far-detuned colliding-probe bi-AM technique. 
	
	\vskip 10pt
	\noindent Under the slowly varying envelope approximation, the Maxwell equations describing the evolution of both probe fields $\mathbf{E}_{a}$ in the moving frame ($\xi=z-ct$, $\eta=z$) are 
	\begin{equation}
	\left(\frac{\partial\Omega_{mn}}{\partial \eta}\right)=i\kappa_{nm}\rho_{mn}, \quad(m=2,4;n=1,3)
	\end{equation}
	where $\kappa_{nm}={\cal N}_0\omega_a|D_{nm}|^2/(\hbar c)$ with $\omega_a$ and ${\cal N}_0$ being the field frequency and atom number density, respectively. Neglecting linear absorption we get for $\Omega_{p1}^{(+)}$
	\begin{equation}
	\frac{\partial\Omega_{p1}^{(+)}}{\partial\eta}\approx\frac{i\alpha_{p1}(\!1+id_{p1})S_{p1}^{(-)}}{(1+d_{p1}^2\!)(-d_B+i)}\left\{\Omega_{p1}^{(+)}+\frac{{\Omega_{p2}^{(+)}}^*\Omega_{p2}^{(-)}}{{\Omega_{p1}^{(-)}}^*}\left(\frac{1+d_{p1}^2}{1\!+d_{p2}^2}\right)\right\},
	\end{equation}
	where $\alpha_{p1}=\kappa_{23}/\Gamma(1+d_{p1}^2)$. Equation (4) contains four coupled nonlinear wave equations, similar to Eq. (5), for four complex field components. We now show that the physics of giant enhancements to NMORE signal and magnetic field sensitivity by the inelastic wave-scattering process can be understood by examining these wave equations in photon number formalism. 
	
	\vskip 10pt
	\noindent
	Letting $\Omega_{p1}^{(\pm)}=R_{\pm}\,e^{i\theta_{\pm}}$, $\Omega_{p2}^{(\pm)}=r_{\pm}\,e^{i\phi_{\pm}}$, where $R_{\pm}$, $\theta_{\pm}$, $r_{\pm}$ and $\phi_{\pm}$ are real quantities, defining $S_{p1}^{(\pm)}=|R_{\pm}|^2/\gamma\Gamma$ and $S_{p2}^{(\pm)}=|r_{\pm}|^2/\gamma\Gamma$, then Eq. (5) yields 
	\begin{subequations}
		\begin{align}
			\frac{\partial S_{p1}^{(\pm)}}{2\partial\eta}&\approx\alpha_{p1}\frac{(1\pm\!d_{p1}d_B)S_{p1}^{(-)}S_{p1}^{(+)}}{(1+d_{p1}^2)(1+d_B^2)}\left\{1+G\,{\rm Cos}(2\theta_0)\right\},\\
			\frac{\partial\theta_{\pm}}{\partial\eta}
			&\approx\alpha_{p1}\frac{(d_{p1}\mp d_B)S_{p1}^{(\mp)}}{(1+d_{p1}^2)(1+d_B^2)}\left\{1+G\,{\rm Cos}(2\theta_0)\right\}.
		\end{align}
	\end{subequations}
	where the gain function $G=(1+d_{p1}^2)\sqrt{S_{p2}^{(+)}S_{p2}^{(-)}}${\Large /}$(1+d_{p2}^2)\sqrt{S_{p1}^{(+)}S_{p1}^{(-)}}$ and $\theta_0$ is the polarization cross angle between the two linearly-polarized probe fields. When $|d_{p1}|\!>\!|d_{p2}|$, we have $G\gg 1$. The Maxwell equations for 
	$S_{p2}^{(\pm)}$ and $\phi_{\pm}$ can be similarly obtained.
	
	\vskip 10pt
	\noindent Letting $\Delta S_{p1}\!=\!S_{p1}^{(+)}\!-\!S_{p1}^{(-)}$, $\Delta\theta\!=\!\theta_{+}\!-\!\theta_{-}$, and $\theta_0=0$ we immediately obtain
	(cp: colliding-probe, sp: single-probe)
	\begin{subequations}
		\begin{align}
			\left(\frac{\partial\Delta S_{p1}}{\partial\eta}\right)_{\rm cp}&\approx
			\frac{4\alpha_1d_{p1}d_{B}S_{p1}^{(-)}S_{p1}^{(+)}}{(1+d_{p1}^2)(1+d_{B}^2)}(1+G)\gg\left(\frac{\partial\Delta S_{p1}}{\partial\eta}\right)_{\rm sp}\approx 0,\\
			\left(\frac{\partial\Delta\theta}{\partial\eta}\right)_{\rm cp}&
			\approx\Theta_{\rm sp}
			\left\{1+\left(1+\frac{S_{p1}^{(+)}}{S_{p1}^{(-)}}\right)G\right\}>\Theta_{\rm sp},
		\end{align}
	\end{subequations}
	where $\Theta_{\rm sp}$ is exactly the single-probe NMORE per unit length with power broadening neglected \cite{r20a}. 
	
	\vskip 10pt
	\noindent 
	\section{Discussion}
	Remarkably, all physical effects of a single-probe AM as well as the colliding-probe bi-AM can be qualitatively understood and predicted by Eqs. (6) and (7). In fact, Eqs. (7a) and (7b) separately give the colliding-probe bi-AM NMORE heterodyne signal enhancement and theoretical detection sensitivity limit.
	\begin{figure}[htb]
		\centering
		\includegraphics[width=14 cm]{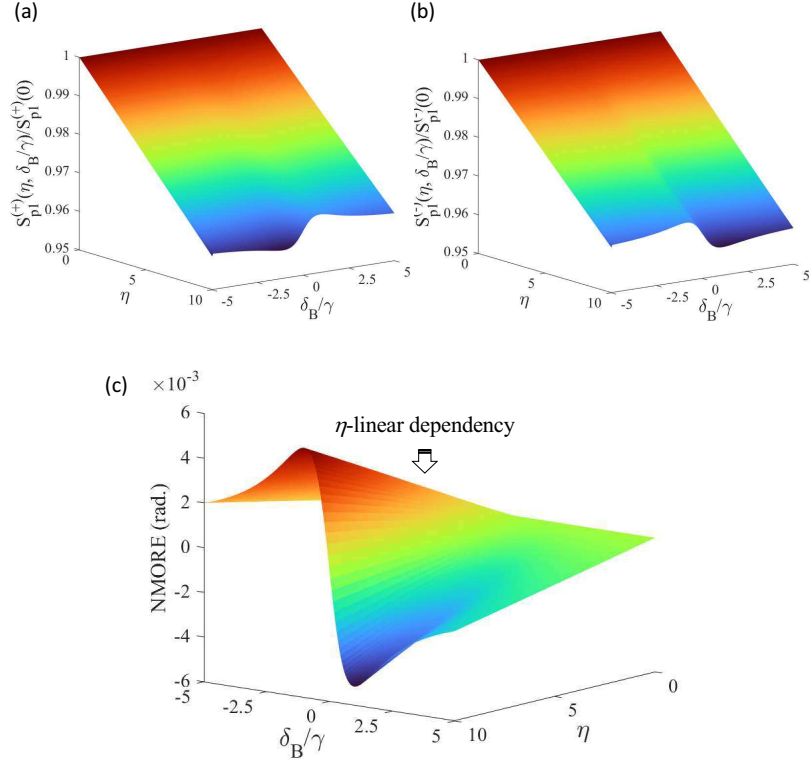}
		\caption{{\bf Single-probe AM NMORE blockade and the resulting linear-growth.} (a) and (b) Numerical calculations of the field intensities $S_{+}$ and $S_{-}$ as functions of propagation distance and magnetic field. (c) Symmetry-forced NMORE blockade strongly suppresses nonlinear NMORE growth, resulting in very weak linear growth. Parameters: $S_{\pm}{(0)}=25$, $d_{p}=10$, $\alpha_1=\alpha_2=0.0065$. }
	\end{figure}
	
	\vskip 5pt
	\noindent {\bf 3.1. Single-probe NMORE blockade.} It is instructive at this moment to look the case where only one probe field $\mathbf{E}_{p1}$ is present. For this widely studied single-probe $F=1$ scheme we neglect the third terms in Eqs. (6a) and (6b) (i.e., let $\mathbf{E}_{p2}=0$). Since the probe is the only energy source both of its components must add up, at any given propagation distance $\eta$, to enforce $S_-=S_0-S_+$, where $S_0=\Omega_0^2/\gamma\Gamma$ is the total probe field energy at $\eta$. Consequently, Eqs. (6a) and (6b) become
	\begin{subequations}
		\begin{align}
			&\frac{dS_+}{d\eta}\approx-\alpha S_{+}+2\alpha\frac{(1+d_{p1}d_{B})(S_0-S_{+})S_{+}}{(1+d_{p1}^2)(1+d_{B}^2)},\\
			&\frac{d\theta_+}{d\eta}=-\alpha d_{p1}\left\{1-\frac{(S_0-S_+)\left(1-\frac{d_{B}}{d_{p1}}\right)}{(1+d_{p1}^2)(1+d_{B}^2)}\right\}.
		\end{align}
	\end{subequations}
	We immediately notice the gain clamping effect, $(S_0-S_+)S_+$, in Eq. (8a). Integrating Eq. (8a) over the propagation distance, we obtain
	\begin{equation}
	S_+(\eta)=\frac{S_+(0)\,e^{-\alpha(1-{\cal A}S_0)\eta}}{1+\left(\frac{{\cal A}S_+(0)}{1-{\cal A}S_0}\right)[1-e^{-\alpha(1-{\cal A}S_0)\eta}]}
	\approx S_+(0)=\frac{S_0}{2}\rightarrow\left(\frac{\partial\Delta S}{\partial\eta}\right)_{\rm sp}\approx 0,
	\end{equation}	
	where ${\cal A}=2(1+d_{p1}d_{B})/(1+d_{p1}^2)(1+d_{B}^2)\ll 1$ and $\Delta S\!=\!S_{+}\!-\!S_{-}$. Equation (9) is exactly the conclusion stated before, i.e., no appreciable change is allowed for each field component. {\bf This is a single-probe energy-symmetry-based self-limiting effect, an intrinsic symmetry-enforced NMORE blockade that blocks any directional energy flow}. Figures 3(a) and 3(b) show numerical solutions of $S_{\pm}$ obtained from Eq.(8a). Only $4\%$ field intensity change is allowed by this strong NMORE blockade, as expected. We note that these results agree well with a first-principle numerical calculation without any approximation, a testimonial of the accuracy of the colliding-probe theory. 
	
	\vskip 10pt
	\noindent Inserting Eq. (9) into Eq. (8b) we immediately obtain the magnetic field dependent polarization angular rotation of a single-probe AM with power broadening neglected \cite{r22}, i.e.,
	\begin{equation}
	\Theta_{\rm sp}(\eta)=-\frac{\alpha d_B S_0}{(1+d_{p1}^2)(1+d_B^2)}\eta.
	\end{equation}
	Equation (10) shows very small NMORE effect [see Fig. 3(c)]. If the one-photon detuning is reduced to improve the polarization rotation, the linear absorption will further reduce the NMORE signal amplitude attainable.
	
	\vskip 10pt
	\noindent We stress that both Eqs. (9) and (10) do not exhibit any nonlinear propagation dependency which usually is a key characteristic for a process being associate to a {\bf nonlinear} effect in nonlinear optics. In fact, even cross-phase modulation processes in nonlinear optics, processes that exhibit a cross-component optical power dependency similar to single-probe NMORE, do show nonlinear propagation distance dependency. Indeed, {\bf the weak $\eta-$linear dependency in Eq. (10) is the direct consequence of the NMORE blockade} Eq. (9).
	
	\begin{figure}[htb]
		\centering
		\includegraphics[width=14 cm]{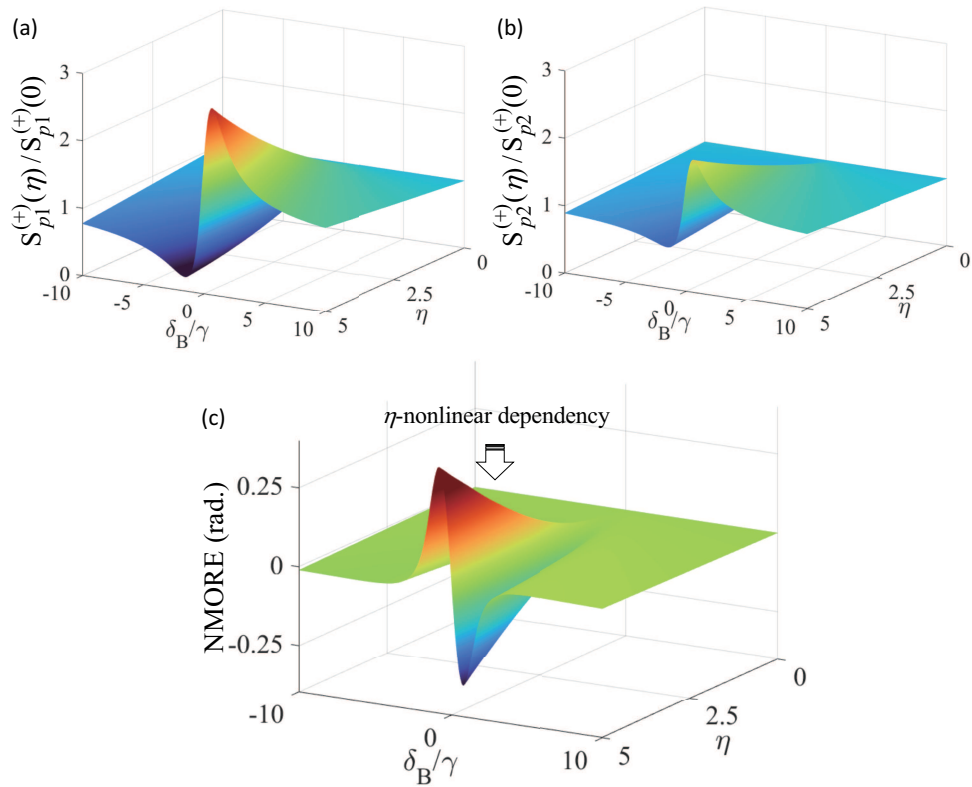}
		\caption{{\bf Highly-nonlinear propagation growth in colliding-probe AM by directional energy flow.} (a) and (b) Field intensities $S_{p1}^{(+)}$ and $S_{p2}^{(+)}$ as functions of propagation distance and magnetic field. Notice that $S_{p2}^{(+)}$ exhibits an``in-phase" growth as $S_{p1}^{(+)}$, indicating an energy circulation (the linear-scaled color code is from 0.5 to 3.0). (c) Colliding-probe bi-AM breaks the NMORE blockade and enables a directional energy flow, resulting in a highly-nonlinear propagation growth of NMORE. Parameters: $S_{p1}^{(\pm)}{(0)}=S_{p2}^{(\pm)}{(0)}=25$, $d_{p1}=10$, $d_{p2}=5$, ${\rm Cos}(2\theta_0)=1$, $\alpha_1=\alpha_2=0.0065$. The linear-scaled color code is from $-$0.3 to 0.3. }
	\end{figure}
	
	\vskip 10pt
	\noindent
	{\bf 3.2. Colliding-probe bi-AM: Breaking the single-probe NMORE blockade.} When the second probe field is introduced the gain function $G$ leads to enhancements to NMORE signal and magnetic field sensitivity. Intuitively, the second, counter-propagating field adds energy to the ground states shared by both probes, lifting the energy-limited growth restriction imposed by the symmetry of the first probe field to itself. More importantly, this initiates a magnetic-field-dependent {\bf directional circulating energy flow} [see Fig. 2(a)], resulting in $\Omega_{p1}^{(\pm)}$ being substantially different from their initial values by propagation. We note that Eq. (6a) describes an inelastic wave scattering process where through {\bf shared and populated intermediate states} a directional energy flow can occur from one probe branch into the other branch depending on branch excitation rates and magnetic field induced level shifting. Generally, the signal in the energy-accepting branch can be substantially larger than a single-probe AM operating under the same conditions. We emphasize that even though $G$ contains multiple fields and must be self-consistently evaluated with all fields, the above qualitative arguments are generally correct.  Experimentally, by adjusting the ratio of two one-photon detunings we can achieve a large increase in NMORE signal interchangeably in either probe channel. 
	
	\vskip 10pt
	\noindent Figures 4(a) and 4(b) show nonlinear growth of field intensities $S_{p1}^{(+)}$ and $S_{p2}^{(+)}$ as functions of propagation distance and magnetic field. Note that field $S_{p2}^{(+)}$ exhibits a similar ``in-phase" growth as $S_{p1}^{(+)}$, indicating ``simultaneous" stimulated emission increase in $|2\rangle\rightarrow|1\rangle$ and $|4\rangle\rightarrow|3\rangle$ transitions, and therefore the energy circulation as depicted in Fig. 2(a) and also qualitatively argued above. 
	
	\vskip 10pt
	\noindent The magnetic field dependent energy circulation can be further shown mathematically by inspecting Eq. (6a) and the Maxwell equations for $S_{p2}^{(+)}$ that couples $|4\rangle\rightarrow|3\rangle$ transition because of its counter-propagating arrangement. In order to establish an energy circulation therefore break the NMORE blockade both of these two field components must have propagation gains (i.e., increased stimulated emissions) for a given magnetic field. This is exactly what the inelastic scattering theory demonstrates. Indeed, the relevant nonlinear contributions can be expressed as
	\begin{subequations}
		\begin{align}
			&\frac{\partial S_{p1}^{(+)}}{2\partial\eta}\approx\alpha_1\frac{ d_{p1}d_{B}}{(1+d_{p2}^2)(1+d_{B}^2)}S_{p1}^{(+)}S_{p1}^{(-)}G(\eta)\\,
			&\frac{\partial S_{p2}^{(+)}}{2\partial(-\eta)}\approx\alpha_1\frac{-d_{p2}d_{B}}{(1+d_{p1}^2)(1+d_{B}^2)}S_{p2}^{(+)}S_{p2}^{(-)}\frac{1}{G(\eta)}.
		\end{align}
	\end{subequations}
	Clearly, both components grow ``in-phase" for the same $d_B$, indicating the energy circulation as discussed above and the other two cross components (i.e., $S_{p1}^{(-)}$ and $S_{p2}^{(-)}$) decrease as propagation. The consequence of this directional energy flow resulting from lifting the NMORE blockade is a large polarization rotation [compare Figs. 4(c) and 3(c)]. To reach a similar signal amplitude without increasing the laser power, the one-photon detuning in a single-probe AM has to be reduced by a factor of 5, an operation condition that inevitably leads to significantly broadened magnetic resonance as well as degradation of the detection sensitivity. When the direction of the magnetic field is reversed the energy circulation is also reversed, as expected. This is exactly what has been observed experimentally.
	\begin{figure}[htb]
		\centering
		\includegraphics[width=14 cm]{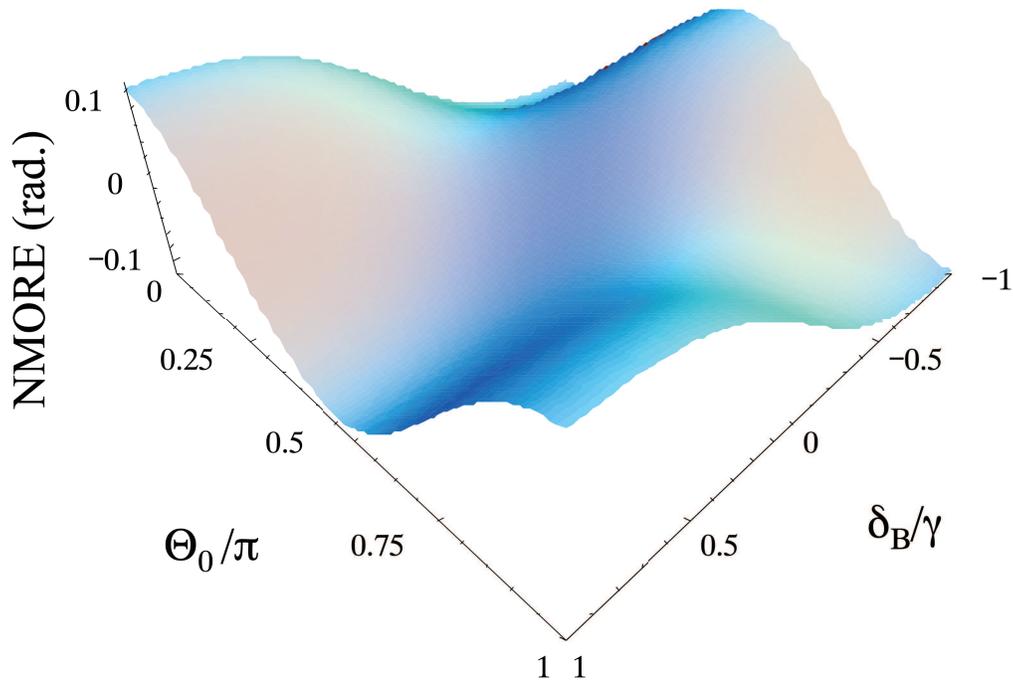}
		\caption{{\bf Colliding-probe bi-AM cross-polarization angular effect.} Numerical calculation using Eq. (7b) shows the $\pi-$periodicity that agrees with the data shown in Fig. 1(b). Parameters: $d_B=\delta_{B}/\gamma=1$ and $\eta=5$. Other parameters see Fig. 4. }
	\end{figure}
	
	\vskip 10pt
	\noindent {\bf 3.3. Enhancement to magnetic field detection sensitivity and NMORE cross-polarization angular dependency.} The magnetic field sensitivity of the colliding-probe bi-AM can be estimated from Eq. (7b) which gives the upper limit of the sensitivity achievable using the new technique. We note that the pre-factor $\Theta_{\rm sp}$ in Eq. (7b) is exactly the single-probe detection sensitivity per unit medium length given in Eq. (10). The $d_{p1}$ dependency of $G$ exactly explains why the detection sensitivity increases as the first probe detuning increases \cite{r22}. Typically, the colliding-probe bi-AM enhanced magnetic field detection sensitivity is about 6-dB greater than the theoretical sensitivity of a single-probe AM operating under the same conditions. Experimentally, even at near resonant conditions (i.e., $d_{p1}\sim 1.5$ Doppler linewidth) where a single-probe AM has reported $\sim fT/\sqrt{Hz}$ sensitivity \cite{r7,r22a}, the colliding-probe technique consistently demonstrates 6-dB or better sensitivity under the same shielding and detector electronics conditions \cite{r22}. Figure 3 shows the NMORE as a function of polarization cross angle $\theta_0$, using Eq. (7b) with ${\rm Cos}(2\theta_0)$ included. The $\pi-$periodicity agrees well with the experimental observations shown in Fig. 1(b). 
	
	\vskip 10pt
	\noindent It is important to stress that the inelastic wave-scattering based correlated colliding-probe bi-AM is fundamental different from the usual four-wave-mixing (FWM) process as well as any electromagnetically induced transparency (EIT) based processes. Typical FWM processes, such as third harmonic generation or double-$\Lambda$ frequency conversion \cite{r21,r16} routinely encountered in nonlinear optics, are elastic processes where the excitation starts from a single fully populated ground state and cycles back to the same ground state. Generally, in each cycle all intermediate states have negligible populations, no appreciable energy storage or population transfer. The ground state Zeeman coherence is negligible. In addition, except the internally generated field all laser fields have no propagation effect other than linear absorption. However, the intensity of the newly generated field grows quadratically as the propagation distance. In the case of parametric FWM generation processes where more than one field are generated the growth of new fields are highly nonlinear with strict angular phase matching requirement (usually forming a radiation cone). The inelastic wave scattering based colliding-probe bi-AM process does not have any of these behaviors. Furthermore, the colliding-probe bi-AM process is not related to any electromagnetically induced transparency (EIT) based processes. In fact, the cross-component optical power dependency shown in all single-probe NMORE processes is contradictory to any EIT-based power dependency. Generally, an EIT process requires a strong coupling field to produce a large ac Stark shift, i.e, an Autler-Townes splitting. Such a strong coupling field preempts any population in the corresponding two-photon terminal state. Furthermore, for any argument attempting to interpret one probe component undergoes an EIT process there is an identically logical counter-argument against it, therefore invalidate any EIT argument. 
	
	\vskip 10pt
	\noindent
	Finally, we also note that because of large one-photon detunings (several GHz), very weak probe fields in $<$nT magnetic field regime the inelastic wave-scattering process cannot be attributed to a pump-probe excitation process or any light-shift-based alignment-orientation effect \cite{r24,r25,r26}. More importantly, a typical pump-probe experiment, regardless of whether it employs saturated absorption spectroscopy or polarization spectroscopy, essentially relies on a two-level atomic system where the pump strongly saturates the transition while the probe is scanned across the absorption profile. In such a case only the {\bf linearized} response of the probe field is considered. The SERF magnetometer, aside from its novel spin-relaxation suppression mechanism, is an example of pump-probe scheme. In fact, the probe field induced angular momentum procession/projection representation widely used in SERF magnetometers with strong on-resonance pumps \cite{r9,r10,r11,r12,r13,r14} is exactly the linearized probe response as in a pump-probe theory \cite{r27}. However, the colliding-probe bi-AM is based on a third-order nonlinear process. Typically, the laser powers are 20 times less than the corresponding one-photon resonance saturation power and effective excitation rates due to large one-photon detunings are two to three orders of magnitude less than the resonance saturation rate, rendering the pump-probe or alignment-orientation arguments invalid.

	\section{Conclusion}
	\vskip 10pt
	\noindent In conclusion, we have shown theoretically and experimentally a colliding-probe bi-AM technique that exhibits significantly large NMORE heterodyne signal enhancement as well as increased magnetic field sensitivity in comparison to widely practiced single-probe AM under the same conditions. These enhancements persist even at near resonance (although a lesser degree) where the single-probe AM has reported $\sim fT/\sqrt{Hz}$ detection sensitivity using sophisticated and complex shielding and electronics. The inelastic wave-scattering based directional energy circulating colliding-probe theoretical framework can qualitatively explain all experimentally observed effects of this new atom magnetometry technique remarkably well.  We emphasize that the far-detuned colliding-probe bi-AM technique significantly reduces probe power induced zero-field shift and broadening, complications that are unavoidable with near-resonance excitation of most single-probe AM techniques. We finally note that the multi-field-dependency of $G$ indicates correlation between the two probe fields. When the choice of detunings is in favor of one probe branch the directional energy flow drains the energy from the other probe branch. As a result the NMORE signal of the energy receiving branch increases significantly and the NMORE signal for the energy contributing branch degrades quickly. This cross-field NMORE SNR correlation is in some way similar to the ``squeezing" effect for a pair of canonical variables. In a sense, the SNRs of two NMOREs from different probe branches may be viewed as a pair of ``squeezing parameters" in this novel correlated colliding-probe bi-atomic magnetometer.
	
\vskip 10pt
\noindent {Acknowledgments}
		LD thanks Dr. B. Liu (SDU) and Dr. F. Zhou (NIST) for representative data, Dr. Changfeng Fang (SDU) and Claire Deng (Thomas Wootton High School) for technical assistance on MATLAB coding and graphics. LD also acknowledges the financial support from SDU.




\begin{thebibliography}{99}
\bibitem{r19} Typically, both probe excitation rates are $<1\%$ of the one-photon saturation rates for the relevant transitions.
\bibitem{r20a} Analytical solutions of single-probe NMORE for three and four-state systems with both light shift and resonance broadening included will be published separately.
\bibitem{r22a} Note that Ref.\cite{r7} uses sophisticated magnetic shielding and RF modulation technique. Our setup has a simple two-layer shield and no fast lock-in amplifier. This shows the great potential of the colliding-probe bi-AM technique.
\bibitem{r25} For small one-photon detunings (such as those in Ref.[3]) both field-dependent frequency shift and power broaden- ing are significant and one must also include strong total field attenuation.
\bibitem{r26} Giant NMORE enhancement reported here has been observed at $<1-nT$ magnetic field regime with both lasers detuned 3-5 GHz from relevant one-photon resonances using only 500 nW/1$\mu$W power for probe $\mathbf{E}_{p1}$ and probe $\mathbf{E}_{p1}$, respectively. The corresponding light shift are only about 10-20 Hz, and therefore cannot be attributed to alignment-to-orientation conversion.
\end{thebibliography}
\end{document}